\documentclass[11pt]{article}
\usepackage[nosort]{cite}
\usepackage{amsmath,amsfonts,amssymb}
\usepackage[small,bf,hang]{caption}
\usepackage{slashed}
\usepackage{latexsym,epsfig}

\def\hybrid{
        \topmargin -20pt
        \oddsidemargin 0pt
        \headheight 0pt \headsep 0pt
        \textwidth 6.25in 
        \textheight 9.5in 
        \marginparwidth .875in
        \parskip 5pt plus 1pt \jot = 1.5ex}

\hybrid

\def\moth{\mathsurround=0pt}
\newdimen\zo \zo=0pt

\def\tick{\leaders\hrule height 0.5ex depth 0pt \hskip 0.5pt}
\def\upboxfill{$\moth \setbox\zo\hbox{\tick}%
  \hskip 3pt\hbox to 0pt{$\tick$\hss}\hrulefill \hbox to 7.5pt{$\tick$\hss}$}

\def\dtick{\leaders\hrule height .34pt depth 0.5ex \hskip 0.5pt}
\def\downboxfill{$\moth \setbox\zo\hbox{\dtick}%
  \hskip 2pt\hbox to 0pt{$\dtick$\hss}\hrulefill \hbox to 2pt{$\dtick$\hss}$}

\def\bec{\begin{center}}
\def\ec{\end{center}}

\def\cB{{\cal B}}

\def\cK{{\cal K}}

\def\cA{{\cal A}}

\def\cA{{\cal A}}

 \def\det{{\rm det\,}}
\def\be{\begin{equation}}
\def\ee{\end{equation}}
\def\bea{\begin{eqnarray}}
\def\eea{\end{eqnarray}}
\def\ba{\begin{array}}
\def\ea{\end{array}}

\def\ie{\rm i.e.\ }
\def\R{{\mathbb R}}

\begin{document}

\thispagestyle{empty}

\begin{flushright}\small
MI-TH-1536
\end{flushright}


\bigskip
\bigskip

\vskip 10mm

\begin{center}

  {\Large{\bf Consistent Pauli reduction on group manifolds}}

\end{center}


\vskip 6mm

\begin{center}

{\bf A. Baguet$^\ast$, C.N. Pope$^{\dag,\ddag}$, H. Samtleben$^\ast$}
\vskip 4mm

$^\ast$\,{\em Universit\'e de Lyon, Laboratoire de Physique, UMR 5672, CNRS et ENS de Lyon,\\
46 all\'ee d'Italie, F-69364 Lyon CEDEX 07, France} \\
\vskip 4mm

$^\dag$\,{\em George P. and Cynthia W. Mitchell Institute \\for Fundamental
Physics and Astronomy \\
Texas A\&M University, College Station, TX 77843-4242, USA}\\

\vskip 4mm
$^\ddag$\,{\em DAMTP, Centre for Mathematical Sciences,
 Cambridge University,\\  Wilberforce Road, Cambridge CB3 OWA, UK}

\end{center}

\vskip1.2cm

\begin{center} {\bf Abstract } \end{center}

\begin{quotation}\noindent

We prove an old conjecture by Duff, Nilsson, Pope and Warner
asserting that the NS-NS sector of supergravity (and more general the bosonic string)
allows for a consistent Pauli reduction on any $d$-dimensional group manifold $G$, 
keeping the full set of gauge bosons of the ${G}\times{G}$ isometry group of the bi-invariant metric on $G$.
The main tool of the construction is a particular generalised Scherk-Schwarz reduction ansatz in double field theory
which we explicitly construct in terms of the group's Killing vectors.
Examples include the consistent reduction from ten dimensions
on $S^3\times S^3$ and on similar product spaces.
The construction is another example of 
globally geometric non-toroidal compactifications
inducing non-geometric fluxes.

\end{quotation}


\section{Introduction}

   Although the idea of Kaluza-Klein theories originated in the
1920s \cite{Kaluza:1921tu,Klein:1926tv}, it was with the advent of higher-dimensional 
supergravities and string theory that the need for developing schemes for 
obtaining
lower-dimensional theories by means of dimensional reduction became
compelling.  The original idea of Kaluza \cite{Kaluza:1921tu}, subsequently 
developed by
Klein \cite{Klein:1926tv}, was straightforwardly extended from a circle reduction
to a reduction on a $d$-dimensional torus.  By this means, for example,
four-dimensional ungauged ${\cal N}=8$ supergravity was constructed, 
by reducing eleven-dimensional supergravity on a 7-torus \cite{Cremmer:1978ds,Cremmer:1979up}. 
A key feature in this, and most other, dimensional reductions is that
one truncates the infinite ``Kaluza Klein towers'' of lower-dimensional
fields that result from the generalised Fourier expansions of the
higher-dimensional fields to just a finite subset, typically, but not
always, just the massless fields.  

    Since the reduction is being applied
to a highly non-linear theory, the question then arises as to whether the
truncation to a finite subset of the fields is a consistent one. One
way to formulate the question is whether in the full  
lower-dimensional theory, prior to the truncation, the equations
of motion of the fields to be truncated are satisfied when one sets
these fields to zero.  The potential danger is that non-linear products
of the fields that are being retained might act as sources for the
fields that are to be truncated.  

   In the case of a circle or toroidal reduction, the consistency of the
truncation is guaranteed by a simple group-theoretic argument.  The
fields that are retained are all the singlets under the $U(1)^d$ isometry
of the $d$-torus, while all the fields that are set to zero are
non-singlets (\ie they are charged under the $U(1)$ factors).  It is
evident, by charge conservation, that no powers of neutral fields can act 
as sources for charged fields, and so the consistency is guaranteed.

  A more general class of dimensional reductions was described by
DeWitt in 1963 \cite{DeWitt:1963}.  In these, 
one takes the internal $d$-dimensional
space to be a compact group manifold $G$, equipped with its bi-invariant
metric.  The isometry group of this metric is $G_L\times G_R$, where $G_L$
denotes the left action of the group $G$ and $G_R$ denotes the right action. 
If all the towers of lower-dimensional fields were retained in a
reduction on the group manifold $G$, then the massless sector would include
the Yang-Mills gauge bosons of the isometry group, $G_L\times G_R$.  
However, in the DeWitt reduction only the gauge bosons of $G_R$ (or,
equivalently and alternatively, $G_L$) are retained.  To be precise,
the lower-dimensional fields that are retained in the truncation are
all those that are singlets under $G_L$.  There is now again a 
simple group-theoretic argument that demonstrates the consistency of the
DeWitt reduction: The fields that are being truncated are all those that
are non-singlets under $G_L$.  It is evident that no non-linear powers
of the $G_L$-singlets that are retained can act as sources for the
fields that are being set to zero, and so the truncation must be 
consistent.
 
  A much more subtle situation arises if one tries to make more general
kinds of dimensional reduction that are not of the toroidal or DeWitt type.
One of the earliest, and most important, examples is the 7-sphere
reduction of eleven-dimensional supergravity.  The massless sector
of the reduced four-dimensional theory contains the fields of 
maximal ${\cal N}=8$ gauged $SO(8)$ supergravity \cite{Duff:1983gq,Biran:1982eg},
but there is no obvious reason why it should be consistent to set
the massive towers of fields to zero.  In particular, one can easily see
that if a generic theory is reduced on $S^7$ (or indeed any other sphere),
then a quadratic product formed from the $SO(8)$ gauge bosons will act as
a source for certain massive spin-2 fields.  This sets off a chain 
reaction that then requires an infinity of four-dimensional 
fields to be retained.  A first indication that something remarkable
might be occurring in the case of eleven-dimensional supergravity and
the $S^7$ reduction was found in \cite{Duff:1984hn}, where it was shown 
that a conspiracy between contributions in the reduction ansatz for the
eleven-dimensional metric and the 4-form field strength resulted in an
exact cancellation of the potentially-troublesome source term for the
massive spin-2 fields that was mentioned above.  Subsequent work
by de Wit and Nicolai in the 1980s \cite{deWit:1986iy}, with more 
recent refinements \cite{deWit:2013ija,Nicolai:2011cy,Godazgar:2013pfa}, has established that the 
truncation to the massless ${\cal N}=8$ gauged $SO(8)$ supergravity is
indeed consistent.  There are a few other examples of supergravity
sphere reductions that also admit analogous remarkable consistent truncations.

   Dimensional reductions on a $d$-dimensional  internal manifold $M_d$
with isometry group
$G$ that admit a consistent truncation to a finite set of fields that
includes all the gauge bosons of the Yang-Mills group $G$ were called 
Pauli reductions in \cite{Cvetic:2003jy}.  (The idea of such reductions
was first proposed, but not successfully implemented, by Pauli in 1953 
\cite{pauli,Straumann:2000zc,ORaifeartaigh:1998pk}.)  It was also observed in \cite{Cvetic:2003jy} 
that in addition
to the necessary condition for consistency that was first seen in 
\cite{Duff:1984hn}, which was essentially the absence of a cubic coupling
of two gauge bosons to the massive spin-2 modes in the untruncated
lower-dimensional theory, a rather different necessary condition
of group-theoretic origin could also be given.  Namely, one can consider
first the (trivially consistent) truncation of the theory when reduced
instead on the torus $T^d$.  The resulting lower-dimensional theory will
have a (non-compact) group $S$ of global symmetries, with a maximal
compact subgroup $K$.    If the higher-dimensional theory were to
admit a consistent Pauli reduction on the manifold $M_d$ then it must be
possible to obtain that theory, with its Yang-Mills gauge group $G$, by 
gauging the theory obtained instead in the $T^d$ reduction.  (Conversely,
by scaling the size of the $M_d$ reduction manifold to infinity, the
gauged theory should limit to the ungauged one.)  This will only be possible
if the isometry group $G$ of the manifold $M_d$ is a subgroup of the
maximal compact subgroup $K$ of the global symmetry group $S$ of the
$T^d$ reduction. 

   A generic theory will not satisfy the above necessary condition for 
admitting a consistent Pauli reduction.  For example, pure
Einstein gravity in $(n+d)$ dimensions gives rise, after reduction on
$T^d$, to an $n$-dimensional theory with $S=GL(d,\R)$ global symmetry,
whose maximal compact subgroup is $K=SO(d)$.  By contrast, the
isometry group of the $d$-sphere is $G=SO(d+1)$, which is thus not
contained within $K$.   The situation is very different if we 
consider certain supergravity theories, such as eleven-dimensional 
supergravity.  If it is reduced on $T^7$ the resulting four-dimensional
ungauged theory has an enhanced $E_{7(7)}$ global symmetry, for which
the maximal compact subgroup is $K=SU(8)$.  This is large enough to
contain the $G=SO(8)$ isometry group of the 7-sphere, and thus this
necessary condition for consistency of the truncation in the $S^7$ reduction
is satisfied.

  It is evident from the above discussion that if an $(n+d)$-dimensional
theory is to admit a consistent Pauli reduction on $S^d$, in which all
the Yang-Mills gauge bosons of the isometry group $SO(d+1)$ are retained,
then the theory must have some special features that lead to its 
$T^d$ reduction yielding a massless truncation with some appropriate
enhancement of the generic $GL(d,\R)$ global symmetry group.  Similarly,
one may be able to rule out other putative consistent Pauli reductions
by analogous arguments.

   This brings us to the topic of the present paper.  It was observed
in \cite{Duff:1986ya} that in a reduction of the $(n+d)$-dimensional
bosonic string on a group manifold $G$ of dimension $d$, the 
potentially dangerous trilinear coupling of a massive spin-2 mode to
bilinears built from the Yang-Mills gauge bosons of $G_L\times G_R$ was in
fact absent.  
On that basis, it was conjectured in \cite{Duff:1986ya} that there
exists a consistent Pauli reduction of the $(n+d)$-dimensional
bosonic string on a group manifold $G$ of dimension $d$, yielding a
theory in $n$ dimensions containing the metric, the Yang-Mills gauge
bosons of $G_L\times G_R$, and $d^2+1$ scalar fields which parameterise
$\R\times SO(d,d)/(SO(d)\times SO(d))$.  Further support for
the conjectured consistency was provided in \cite{Cvetic:2003jy}, where it
was observed that the $K=SO(d)\times SO(d)$ maximal compact subgroup
of the enhanced $O(d,d)$ global symmetry of the $T^d$
reduction of the bosonic string is large enough to contain the
$G_L\times G_R$ gauge group as a subgroup.  

In this paper, we shall present a complete and constructive 
proof of the consistency of the Pauli reduction of the bosonic
string on the group manifold $G$. Our construction makes use of 
the recent developments realising non-toroidal compactifications 
of supergravity via generalised Scherk-Schwarz-type reductions~\cite{Scherk:1979zr} 
on an extended spacetime within duality covariant reformulations 
of the higher-dimensional supergravity theories~\cite{Aldazabal:2011nj,Geissbuhler:2011mx,Berman:2012uy,Aldazabal:2013mya,Lee:2014mla,Hassler:2014sba,Hohm:2014qga,Cho:2015lha}. 
In this language, consistency of a truncation ansatz translates into
a set of differential equations to be satisfied by the group-valued Scherk-Schwarz twist matrix $U$
encoding all dependence on the internal coordinates.
Most recently, this has been put to work 
in the framework of exceptional field theory in order to derive the 
full Kaluza-Klein truncation of IIB supergravity on a 5-sphere 
to massless ${\cal N}=8$ supergravity in five dimensions~\cite{Baguet:2015xha,Baguet:2015sma}.
In this paper, we explicitly construct the $SO(d,d)$ valued twist matrix
describing the Pauli reduction of the bosonic string on a group manifold $G$
in terms of the Killing vectors of the group manifold. We show that it satisfies
the relevant consistency equations, 
thereby establishing consistency of the truncation.
From the Scherk-Schwarz reduction formulas we then read off the explicit 
truncation ans\"atze  for all fields of the bosonic string. We find agreement
with the linearised ansatz proposed in~\cite{Duff:1986ya} and for the metric
we confirm the non-linear reduction ansatz conjectured in \cite{Cvetic:2003jy}.

Our solution for the twist matrix straightforwardly generalises to the case when $G$ is a
non-compact group. In this case, the construction describes the consistent reduction of the
bosonic string on an the internal manifold $M_d$ whose isometry group is given
by  the maximally compact subgroup $K_{L}\times K_{R}\subset G_L\times G_R$.
The truncation retains not only the gauge bosons of the isometry group, but
the gauge group of the lower-dimensional theory enhances to the full non-compact $G_L\times G_R$.
At the scalar origin, the gauge group is broken down to its compact part.
This is a standard scenario in supergravity with non-compact gauge groups:
for the known sphere reductions the analogous generalisations describe the
compactification on hyperboloids $H^{p,q}$ and lower-dimensional theories with $SO(p,q)$
gauge groups~\cite{Hull:1988jw,Cvetic:2004km,Hohm:2014qga,Baron:2014bya}.

The paper is organised as follows. In section~2 we briefly review the 
$O(d,d)$ covariant formulation of the low-energy effective action of the 
$(n+d)$-dimensional bosonic string.
In section~3 we review how this framework allows the reformulation of
consistent truncations of the original theory as generalised Scherk-Schwarz 
reductions on the extended space-time. We spell out the consistency equations
for the Scherk-Schwarz twist matrix and construct an explicit solution in terms
of the Killing vectors of the bi-invariant metric on a $d$-dimensional group manifold $G$.
For compact $G$, the construction results in the Pauli reduction of the bosonic
string on $G$ to a lower-dimensional theory with gauge group $G\times G$. 
For non-compact $G$, the construction gives rise to a consistent truncation 
on an internal space $M_d$ whose isometry group is given by two copies of the 
maximally compact subgroup $K\subset G$. Again, the gauge group
of the lower-dimensional theory is $G\times G$.
In section~4, we work out the complete non-linear 
reduction ansatz for the higher-dimensional fields, i.e.\
metric, two-form and dilaton.
We discuss our findings in section~5, in particular the examples of
consistent truncations of ten-dimensional $N=1$ supergravity down
to four dimensions on products of spheres and hyperboloids.

\section{
$O(d,d)$ covariant formulation of the
$(n+d)$-dimensional bosonic string}

Our starting point is 
the $(n+d)$-dimensional bosonic string
(or NS-NS sector of the superstring)
 \bea
 S &=& \int dX^{n+d}\,\sqrt{|\hat{G}|}\,e^{-2\phi}\Big(  {R}
   +4\,{\hat G}^{\hat\mu\hat\nu}\partial_{\hat\mu}\phi \partial_{\hat\nu}\phi 
    -\frac{1}{12}\,{H}^{\hat\mu\hat\nu\hat\rho}{H}_{\hat\mu\hat\nu\hat\rho}
 \Big)\;,
 \label{string}
 \eea
with dilaton $\phi$ and three-form field strength ${H}_{\hat\mu\hat\nu\hat\rho}\equiv3\,{\partial}_{[\hat\mu} C_{\hat\nu\hat\rho]}$.
As described in the introduction, the conjecture of~\cite{Duff:1986ya} states this theory
admits a consistent Pauli reduction to $n$ dimensions on a $d$-dimensional group manifold $G$ retaining the
full set of $G_L\times G_R$ non-abelian gauge fields,
according to the isometry group of the bi-invariant metric on $G$.
In the following, for the explicit reduction formulas we will use the metric in the Einstein frame
\bea
G_{\hat\mu\hat\nu} &\equiv& e^{-4\beta\phi}\,\hat{G}_{\hat\mu\hat\nu}\;,
\label{framechange} 
\eea
with $\beta=1/(n+d-2)$,
and split coordinates according to
\bea
\{X^{\hat\mu}\}\rightarrow\{x^\mu,y^m\}\;,\qquad
\mu=0, \dots, n-1\;,\quad m=1, \dots, d
\;.
\label{split_dn}
\eea

The key tool in the following construction is
double field theory (DFT)~\cite{Siegel:1993th,Hull:2009mi,Hohm:2010jy,Hohm:2010pp}, 
the duality covariant formulation of the bosonic string. 
Most suited for our purpose, is the reformulation of the action (\ref{string})
in which an $O(d,d)$ subgroup of the full duality group is made manifest~\cite{Hohm:2013nja}.
This is obtained by Kaluza-Klein decomposing all fields according to $n$ external and
$d$ internal dimensions (keeping the dependence on all $(n+d)$ coordinates) and 
rearranging the various components into $O(d,d)$ objects, in terms of which the action
(\ref{string}) can be rewritten in the form
\bea
S &=& \int dx^{n} dY^{2d}\,
 \sqrt{|{\rm g}|}\,e^{-2\Phi}\Big(  \widehat{\cal R}
   +4\,{\rm g}^{\mu\nu}D_{\mu}\Phi D_{\nu}\Phi      -\frac{1}{12}{\cal H}^{\mu\nu\rho}{\cal H}_{\mu\nu\rho}
   +\frac{1}{8}\,{\rm g}^{\mu\nu}D_{\mu}{\cal H}^{MN}D_{\nu}{\cal H}_{MN}
   \nonumber\\
   &&{}\qquad\qquad\qquad\qquad\qquad
   -\frac{1}{4}\,{\cal H}_{MN}{\cal F}^{\mu\nu M}{\cal F}_{\mu\nu}{}^{N}
  + \frac{1}{4}\,{\cal H}^{MN}\partial_M{\rm g}^{\mu\nu}\,\partial_N {\rm g}_{\mu\nu}
    +{\cal R}(\Phi,{\cal H}) 
 \Big)\;.
\label{DFTAction}
\eea
Formally, this theory lives on an extended space of dimension $(n+2d)$ with coordinates $\{x^\mu, Y^M\}$, with
all fields subject to the section constraint $\partial^M \otimes \partial_M \equiv 0$ which effectively removes the $d$
non-physical coordinates. Fundamental $SO(d,d)$ indices $M, N$
are lowered and raised with the $SO(d,d)$ invariant metric $\eta_{MN}$
and its inverse.
Moreover, ${\cal H}_{MN}$ is a symmetric $SO(d,d)$ group matrix, 
${\cal H}_{\mu\nu\rho}$ and ${\cal F}_{\mu\nu}{}^{M}$ are the non-abelian field strengths
of an external two-form ${\cal B}_{\mu\nu}$ and vector ${\cal A}_\mu{}^M$, 
respectively,
and  ${\cal R}(\Phi,{\cal H})$ is the 
scalar DFT curvature~\cite{Hohm:2010pp}.
All derivatives and field strengths in (\ref{DFTAction}) are covariantised
with respect to generalised diffeomorphisms on the extended space. 
Specifically, 
\bea
D_\mu \Phi &=& \partial_\mu \Phi - {\cal A}_\mu{}^M \partial_M \Phi + 
\frac12\,\partial_M {\cal A}_\mu{}^M\;,\nonumber\\
D_\mu {\cal H}_{MN} &=& \partial_\mu {\cal H}_{MN} - {\cal A}_\mu{}^K \partial_K {\cal H}_{MN}
-2\,\partial_{(M} {\cal A}_\mu{}^K {\cal H}_{N)K}+2\,\partial^K {\cal A}_{\mu\,(M}  {\cal H}_{N)K}\;,\nonumber\\
  {\cal F}_{\mu\nu}{}^{M}&=& \partial_{\mu}{\cal A}_{\nu}{}^{M}-\partial_{\nu}{\cal A}_{\mu}{}^{M}
  -\big[{\cal A}_{\mu},{\cal A}_{\nu}\big]_{C}^{M}-\partial^{M}{\cal B}_{\mu\nu}\;, 
 \nonumber\\
   {\cal H}_{\mu\nu\rho}& = & 3\,D_{[\mu}{\cal B}_{\nu\rho]} + 3\,{\cal A}_{[\mu}{}^{N}\partial_{\nu}{\cal A}_{\rho]N}
  -{\cal A}_{[\mu N}\big[{\cal A}_{\nu},{\cal A}_{\rho]}\big]^N_{ C}\;,
\eea
in terms of the Courant bracket $[\cdot,\cdot]_{C}$, see~\cite{Hohm:2013nja} for details.

The section constraint $\partial^M \otimes \partial_M \equiv 0$ is solved by splitting the 
internal coordinates according to
\bea
\{Y^M\}\rightarrow\{y^m, y_m\}
\;,
\label{Ymm}
\eea
in a light-cone basis where
\bea
\eta_{MN} &\equiv& 
\left(
\begin{array}{cc}
0& \delta_{m}{}^{n} \\
\delta^{m}{}_{n} & 0
\end{array}
\right)\;,
\label{etaMN}
\eea
and restricting the dependence of all fields  to the physical coordinates $y^m$ by imposing $\partial^m \equiv 0$, 
thereby reducing the extended space-time in (\ref{DFTAction}) back to $(n+d)$ dimensions. 
Upon breaking the DFT field content accordingly, and rearranging of fields,
the $O(d,d)$ covariant form (\ref{DFTAction}) then reproduces the bosonic string (\ref{string}).
The precise dictionary can be straightforwardly worked out by 
matching the gauge and diffeomorphism transformations of the various fields. 
For the DFT $p$-forms and metric this yield
\be
\begin{split}
  \cA_{\mu}{}^m &= A_{\mu}{}^m\equiv G^{mn}G_{\mu n}\;,\quad \cA_{\mu\, m}=-\,({C}_{\mu\,m}-A_{\mu}{}^nC_{nm})\;,\\
 \quad {\cB}_{\mu\nu}&=C_{\mu\nu}+2A_{[\mu}{}^m C_{\nu]\,m}+A_{[\mu}{}^mA_{\nu]}{}^n C_{mn}+A_{[\mu}{}^m A_{\nu] m}\;,\\
  {\rm g}_{\mu\nu} &=e^{4\beta\phi}\,(G_{\mu\nu}-A_{\mu}{}^m A_{\nu}{}^n G_{mn})\;.
\end{split}  
\label{dictionary1}
\ee
The dictionary for the DFT scalar fields is most conveniently
obtained by comparing the transformation of the DFT vector fields under generalised external diffeomorphisms
to the transformations in the original theory (\ref{string}) and yields
\bea
{\cal H}^{mn}&=& e^{-4\beta\phi} \,G^{mn}\;, \qquad {\cal H}_m{}^n=e^{-4\beta\phi} \,G^{nk}C_{km}\;,\nonumber\\
{\cal H}_{mn}&=&e^{-4\beta\phi} \,G^{kl}C_{km}C_{ln}+e^{4\beta\phi}\,G_{mn}\;,\nonumber\\
e^{\Phi}&=& e^{\frac{\beta}{\gamma}\phi} \,(\det\,G_{mn})^{-1/4}\;,
\label{dictionary2}
\eea
with $\gamma=\frac{1}{n-2}$. 
With the dictionary (\ref{dictionary1}), (\ref{dictionary2}), and imposing $\partial^m\equiv0$,
the $O(d,d)$ covariant action (\ref{DFTAction}) reduces to the original action (\ref{string})
of the bosonic string.
The reduction ansatz on the other hand will be most compactly formulated in terms
of the $O(d,d)$ objects.


\section{Generalised Scherk-Schwarz ansatz and consistency equations}


An important property of the $O(d,d)$ covariant form of the action (\ref{DFTAction}) is the fact that particular solutions and
truncations of the theory take a much simpler form in terms of the $O(d,d)$ objects ${\cal A}_\mu{}^M$, ${\cal H}_{MN}$, etc.,
as opposed to the original fields of the bosonic string (\ref{string}). In particular, consistent truncations to $n$ dimensions
can be described by a generalised Scherk-Schwarz ansatz in 
which the dependence on the compactified coordinates $Y^M$ is carried by
an $SO(d,d)$ matrix $U_M{}^{A}$ and a scalar function $\rho$,
according to \cite{Aldazabal:2011nj,Geissbuhler:2011mx}\footnote{
Since with (\ref{DFTAction}) we use DFT in its split form with internal and external coordinates,
the reduction ansatz (\ref{ScherkSchwarz}) resembles the corresponding ansatz in exceptional 
field theory \cite{Hohm:2014qga}
for the $p$-forms and metric.}
\bea
{\cal H}_{MN} &=& U_M{}^A(y) M_{AB}(x) U_N{}^B(y)\;,\qquad
e^{\Phi} ~=~ \rho^{(n-2)/2}(y) \,e^{\varphi(x)}\;,\nonumber\\
{\cal A}_\mu{}^M &=& (U^{-1})_A{}^M(y)\,A_\mu{}^A(x)\;,\qquad
{\cal B}_{\mu\nu} ~=~ B_{\mu\nu}(x)\;,\nonumber\\
{\rm g}_{\mu\nu} &=&  e^{4\gamma\varphi(x)}\,{g}_{\mu\nu}(x)
\label{ScherkSchwarz}
\;.
\eea
Here, $A_\mu{}^M$ and $B_{\mu\nu}$ are the gauge vectors and two-form of the reduced theory.
The symmetric $SO(d,d)$ group valued matrix $M_{AB}(x)$ 
can be thought of as parametrizing the coset space $SO(d,d)/(SO(d)\times SO(d))$, and
together with $e^\varphi(x)$ carries the $d^2+1$ scalar
fields of the reduced theory.
The ansatz (\ref{ScherkSchwarz}) describes a consistent truncation of (\ref{DFTAction}), 
provided $U_M{}^A$ and $\rho$ satisfy the consistency equations
\bea
\eta_{D[A}\,(U^{-1})_B{}^M (U^{-1})_{C]}{}^N \partial_M U_N{}^D&=&f_{ABC}~=~{\rm const.}
\;,
\label{consistent1}
\\
\rho^{-1}\,\partial_M \rho &=&  -\gamma\,(U^{-1})_A{}^N\partial_N U_M{}^A
\;,
\label{consistent2}
\eea
with the $SO(d,d)$ invariant constant matrix $\eta_{AB}$ and $\gamma=\frac{1}{n-2}$.
If $U_M{}^A$ and $\rho$ in addition depend only on the physical coordinates on the extended space (\ref{Ymm})
\bea
\partial^m U_M{}^A &=& 0 ~=~ \partial^m \rho
\;,
\label{consistentS}
\eea
the ansatz (\ref{ScherkSchwarz}) likewise describes a consistent truncation of the original theory (\ref{string}).
 As a consequence of this section condition, the Jacobi identity is automatically satisfied for $f_{ABC}$ upon using its explicit expression $\eqref{consistent1}$
\bea
[X_A,X_B]=-X_{AB}{}^CX_C
\label{jacobi}
\eea
where we have introduced the generalised structure constant $X_{AB}{}^C=f_{[ABD]}\eta^{DC}$.
Then, for a given solution of (\ref{consistent1}), (\ref{consistent2}),
the explicit reduction formulas for the original fields are obtained by combining (\ref{ScherkSchwarz})
with the dictionary (\ref{dictionary1}), (\ref{dictionary2}),
as we will work out shortly.
 
In order to explicitly solve the generalised Scherk-Schwarz consistency conditions (\ref{consistent1})--(\ref{consistentS}), let us first note that 
with the index split (\ref{Ymm}), and the parametrization 
\bea
U_M{}^A ~=~ \eta^{AB} \left \{   {\cal Z}_{B\,m},  {\cal K}_B{}^{m}  \right\}
\;,\qquad
(U^{-1})_A{}^M ~=~ \left \{  {\cal K}_A{}^m , {\cal Z}_{A\,m} \right\}
\;,
\label{UI}
\eea
of the $SO(d,d)$ matrix, equation  (\ref{consistent1}) turns into
\bea
{\cal L}_{{\cal K}_A} {\cal K}_B{}^m &=& 
-X_{AB}{}^C\,{\cal K}_C{}^m
\;,\nonumber\\
{\cal L}_{{\cal K}_A} {\cal Z}_{B\,m}
 + {\cal K}_B{}^n\left(  \partial_m {\cal Z}_{A\,n}-\partial_n {\cal Z}_{A\,m} \right)&=& 
-X_{AB}{}^C\,{\cal Z}_{C\,m}
\;.
\label{system}
\eea
The $SO(d,d)$ property of $U_M{}^A$ translates into
\bea
2\,{\cal K}_{(A}{}^m {\cal Z}_{B)\,m} &=& \eta_{AB}
~\equiv~
\left(
\begin{array}{cc}
0& \delta_{a}{}^{b} \\
\delta^{a}{}_{b} & 0
\end{array}
\right)\;.
\label{KZSO}
\eea

In the following, we will construct an explicit solution of (\ref{system}), (\ref{KZSO}) 
in terms of the Killing vectors of the bi-invariant metric on a $d-$dimensional group manifold $G$.
For compact $G$, the resulting reduction describes the Pauli reduction of 
the bosonic
string on $G$.  For non-compact $G$, this describes a consistent 
truncation on an
internal space $M_d$ with isometry group given by two copies of the 
maximally compact subgroup $K\subset G$.
Specifically, we choose the ${\cal K}_A$ as linear combinations of the $G_L\times G_R$ Killing vectors 
$\{L_a^m, R_a^m\}$,
in the following way
\bea
{\cal K}_{A}{}^m &\equiv& \{ L_a{}^m+R_a{}^m, L^{a\,m}-R^{a\,m} \}\;,
\label{Killing}
\eea
with their algebra of Lie derivatives given by
\bea
{\cal L}_{L_a} L_b = -f_{ab}{}^{c}\,L_c\;,\qquad
{\cal L}_{L_a} R_b = 0\;,\qquad
{\cal L}_{R_a} R_b = f_{ab}{}^{c}\,R_c\;,
\label{LRalgebra}
\eea
in terms of the structure constants $f_{ab}{}^c$ of $\mathfrak{g}\equiv {\rm Lie}\,G$,
and with indices $a, b, \dots$, raised and lowered by the associated Cartan-Killing form 
$\kappa_{ab}\equiv f_{ac}{}^d f_{bd}{}^c$.
Moreover, the bi-invariant metric on the group manifold can be expressed by
\bea
\tilde{G}^{mn} &\equiv& -4\,L_a{}^m L^{a\,n} ~=~ -4\,R_a{}^m R^{a\,n}
\;.
\label{Gtilde}
\eea
With (\ref{LRalgebra}), the ansatz (\ref{Killing}) solves the first equation of (\ref{system}), 
with structure constants $X_{AB}{}^C$ given by
\bea
X_{abc} &=& f_{abc} \;,\quad
X_a{}^{bc} ~=~ f_a{}^{bc}\;,\quad
X^a{}_{b}{}^{c} ~=~ f^a{}_{b}{}^{c}\;,\quad
X^{ab}{}_{c} ~=~f^{ab}{}_{c} 
\;,
\label{XABC}
\eea
and all other entries vanishing. Indeed, these structure constants are
of the required form $X_{AB}{}^C = f_{[ABD]}\eta^{DC}$, c.f.~(\ref{jacobi}).
We may define the $G_L\times G_R$ invariant Cartan-Killing form of the algebra (\ref{jacobi})
\bea
\kappa_{AB} &\equiv& \frac12 X_{AC}{}^D X_{BD}{}^C ~=~ 
\left(
\begin{array}{cc}
\kappa_{ab}& 0 \\
0 & \kappa^{ab}
\end{array}
\right)\;,
\label{kappa}
\eea
such that the Killing vectors (\ref{Killing}) satisfy
\bea
\kappa^{AB}\,
{\cal K}_{A}{}^m {\cal K}_{B}{}^n~=~
-\,\tilde{G}^{mn}
\;,
\qquad
\eta^{AB}\,
{\cal K}_{A}{}^m {\cal K}_{B}{}^n &=& 0\;,
\label{defrel}
\eea
and moreover $\kappa^{AB}\eta_{AB}=0$\,.

In order to solve the second equation of (\ref{system}), with the same structure constants (\ref{XABC}),
we start from the ansatz\footnote{
Let us stress that our notation is such that adjoint $G$ indices $a, b, \dots$ are raised and lowered with the
Cartan-Killing form $\kappa_{ab}$, 
whereas fundamental ${SO}(d,d)$ indices $A, B, \dots$ are raised and lowered with the ${SO}(d,d)$ invariant 
metric $\eta_{AB}$ from (\ref{KZSO}) and {\em not} with the $G$-dependent Cartan-Killing form $\kappa_{AB}$ from (\ref{kappa}).}
\bea
{\cal Z}_{A\,m} &=& -\,\kappa_{A}{}^B\,{\cal K}_{B\,m} + {\cal K}_A{}^n\,\tilde{C}_{nm}
\;.
\label{ansatzZ}
\eea
Here, the space-time index in the first term has been lowered with the group metric $\tilde{G}_{mn}$ from (\ref{Gtilde}), 
and $\tilde{C}_{mn}=\tilde{C}_{[mn]}$ represents an antisymmetric 2-form, such that the $SO(d,d)$ property (\ref{KZSO})
is identically satisfied. With this ansatz for ${\cal Z}_{A\,m}$, the second equation of (\ref{system}) turns into
\bea
\kappa_A{}^C {\cal K}_B{}^n \left(  \partial_n {\cal K}_{C\,m} -\partial_m {\cal K}_{C\,n}\right)
-
3\,{\cal K}_A{}^k\, {\cal K}_B{}^n \, \partial_{[k} \tilde{C}_{mn]} &=& 
2\,\eta^{DE}\,
X_{A(E}{}^C\,\kappa_{B)C}\,{\cal K}_{D\,m}
\;.
\label{2E}
\eea
The right-hand side of $\eqref{2E}$ vanishes by invariance of the Cartan-Killing form  $\kappa_{AB}$. 
From $\eqref{defrel}$, one derives the following identity 
\bea
\partial_{[m}\cK_{A\,n]}=X_{AC}{}^B\kappa^{CD}\cK_{B\,m}\cK_{D\,n}\,,
\eea
for the derivative of the Killing vectors. Inserting this relation
in $\eqref{2E}$ gives
\bea
3\,\cK_{A}{}^k\partial_{[k}\tilde{C}_{mn]}=2 \,X_{A}{}^{BC} \cK_{B\, m}\cK_{C\, n}\,,
\label{3E}
\eea 
where we have used $\kappa_A{}^E X_{ED}{}^C \kappa^{DB} = X_{A}{}^{BC}$.
We note that both sides of this equation vanish under projection with $\eta^{DA}\cK_{A\,p}$
as a consequence of (\ref{defrel}). Projecting instead with $\kappa^{DA}\cK_{A\,p}$, 
equation (\ref{3E}) reduces to an equation for $\tilde{C}_{mn}$
\bea
3\,\partial_{[k}\tilde{C}_{mn]}&=& \tilde{H}_{kmn}
~\equiv~
-2 X^{ABD}\kappa_D{}^C\cK_{A\, k}\cK_{B\, m}\cK_{C\, n}
\;.
\label{equC}
\eea
Explicitly, the flux $\tilde{H}_{kmn}$ takes the form
\bea
\tilde{H}_{kmn}
&=&
-16\,f^{abc}\,L_{a\,k} L_{b\,m} L_{c\,n}  ~=~ -16 \,f^{abc}\,R_{a\,k} R_{b\,m} R_{c\,n} ,
\eea
and can be integrated since
$\partial_{[k} \tilde{H}_{lmn]} = 0$,
due to the Jacobi identity on $f_{abc}$\,.
We have thus solved the second equation of (\ref{system}).

With (\ref{Killing}), (\ref{ansatzZ}), the remaining consistency equation (\ref{consistent2}) reduces to
\bea
(n-2)\, \cK_A{}^m\partial_m \log{\rho} &=& \partial_m \cK_A{}^m~=~ -\tilde\Gamma_{mn}{}^{m} \cK_A{}^n \;,\nonumber\\
&&\Longrightarrow\qquad \rho ~=~ (\det{\tilde{G}_{mn}})^{-\gamma/2} \;.
\eea
We have thus determined the $SO(d,d)$ matrix $U_M{}^A$ and the scalar function $\rho$
solving the system (\ref{consistent1}), (\ref{consistent2}) in terms of the  
Killing vectors on a group manifold $G$, and a two-form determined by (\ref{equC}).
The resulting structure constants are given by (\ref{XABC}) such that the 
gauge group of the reduced theory is given by $G_L\times G_R$\,.

\section{Reduction ansatz and reduced theory}

We now have all the ingredients to read off the 
full non-linear reduction ansatz of the bosonic string (\ref{string}).
Combining the DFT reduction formulas (\ref{ScherkSchwarz})
with the dictionary (\ref{dictionary1}), (\ref{dictionary2}),
and the explicit expressions (\ref{Killing}), (\ref{ansatzZ}) for the
Scherk-Schwarz twist matrix, 
we obtain 
\bea
ds^2 &=&
 \Delta^{-2\gamma}(x,y)\,{g}_{\mu\nu}(x) \, dx^\mu dx^\nu  
\nonumber\\
 &&{}
 + 
G_{mn}(x,y)
\left( dy^m + {\cal K}_{A}{}^m(y) A_\mu^{A}(x)  dx^\mu\right)\left( dy^n +{\cal K}_{B}{}^n(y) A_\nu^{B}(x)   dx^\nu\right)
 \;,
 \label{reducG}
\eea
for the metric in the Einstein frame,
with $G_{mn}(x,y)$ given by the inverse of
\bea
G^{mn}(x,y) &=& \Delta^{2\gamma}(x,y)\, \cK_A{}^m(y)\cK_B{}^n(y) e^{4\gamma\varphi(x)} M^{AB}(x)
\;.
\label{Ginv}
\eea
The dilaton and the original two-forms are given by
\bea
\begin{split}
e^{4\beta\phi} &= \Delta^{2\gamma}(x,y) \,e^{4\gamma\varphi(x)}\,\\ 
C_{mn} &= \widetilde{C}_{mn}(y)+\Delta^{2\gamma}(x,y)\,\kappa_A{}^D\cK_{D\,m}\cK_{B}{}^p G_{pn}(x,y) \, e^{4\gamma\varphi(x)} M^{AB}(x)\;,\\
C_{\mu \,m}&=\left(\kappa_A{}^D\cK_{D\,m}+\Delta^{2\gamma}(x,y)\,\kappa_C{}^E\cK_A{}^n\cK_{E\,n}\cK_{D}{}^p G_{pm}(x,y) \, e^{4\gamma\varphi(x)}M^{CD}(x)\right)A_{\mu}{}^A(x)\;,\\
C_{\mu\nu}&=B_{\mu\nu}(x)-\kappa_B{}^C\cK_A{}^m\,\cK_{C\,m}\,A_{[\mu}{}^A(x)A_{\nu]}{}^B(x)\\
&\quad{}-\Delta^{2\gamma}(x,y)\,\kappa_C{}^E\,\cK_{B}{}^n\cK_{E\,n}\,\cK_{D}{}^p\,\cK_{A}{}^{m}\,G_{pm}(x,y)\,e^{4\gamma\varphi(x)} M^{CD}(x)
\,A_{[\mu}{}^A(x)A_{\nu]}{}^B(x)\;.
\end{split}
\label{reducform}
\eea
where we have introduced the function $\Delta^2(x,y)\equiv\det(\tilde{G}_{mn}(y))^{-1}\,\det(G_{mn}(x,y))$.
In these expressions, all space-time indices on the Killing vectors ${\cal K}_A{}^m$ are raised and lowered
with the metric $\tilde{G}_{mn}(y)$ from (\ref{Gtilde}), rather than with the full metric $G_{mn}(x,y)$\,.
For the group manifold $G=SU(2)$, the construction describes the $S^3$ reduction of the bosonic string,
for which the full reduction ansatz has been found in \cite{Cvetic:2000dm}. For general compact groups, the
reduction ansatz for the internal metric (\ref{Ginv}) was correctly conjectured in \cite{Cvetic:2003jy}.\footnote{
The translation uses an explicit parametrization of the $SO(d,d)$ matrix $M_{AB}$ in a basis where 
$\eta_{AB}$ is diagonal, as
\bea
\tilde{M}_{AB} &=& 
\begin{pmatrix}
(1 +  P  P^t)^{1/2}&{P} \\
{P}^t &(1 +  P^t  P)^{1/2}
\end{pmatrix}
\;,
\nonumber
\eea
in terms of an unconstrained $d\times d$ matrix ${P}_a{}^b$\;.
}

In order to compare our formulas to the linearised result given in \cite{Duff:1986ya}, we first note that
for compact $G$, we may normalise the Cartan-Killing form as $\kappa_{AB}=-\delta_{AB}$,
such that the background (at $M_{AB}(x)=\delta_{AB}$) is given by
\bea
\mathring{{G}}_{mn} &=& \tilde{G}_{mn}\;,\qquad \mathring{C}_{mn} ~=~ \tilde{C}_{mn}\;,\qquad
\mathring{\phi} ~=~ 0
\;.
\label{background_cp}
\eea
We then linearise the reduction
formulas (\ref{reducG})--(\ref{reducform}) around the scalar origin
\bea
M_{AB}(x)&=\delta_{AB}+m_{AB}(x) + \dots
\;,
\label{flucM}
\eea
and (back in the string frame) obtain
\bea
\begin{split}
\hat{G}_{mn}(x,y)&=\tilde{G}_{mn}(y)+\hat{h}_{mn}(x,y)+\dots \,,\quad 
{C}_{mn}(x,y)=\tilde{C}_{mn}(y)+\hat{k}_{mn}(x,y) + \dots \,,\\
\end{split}
\eea
with
\bea
\begin{split}
\hat{h}_{mn}(x,y) &= -m_{AB}(x)\,\cK^A{}_m(y)\,\cK^B{}_n(y)\,,\\
\hat{k}_{mn}(x,y) &= m_{AB}(x)\,\kappa^{AD}\cK_{D\,m}(y)\,\cK^{B}{}_n(y)\,,
\end{split}
\label{fluct}
\eea
as well as
\bea
\phi &=\varphi(x)+\frac14\tilde{G}^{mn}\hat{h}_{mn}+\dots\,,
\eea
for the dilaton, where we have used the linearisation $\Delta(x,y)=1+\frac12\tilde{G}^{mn}\hat{h}_{mn}-2d\beta \phi+\dots$. 
Parametrizing the scalar fluctuations (\ref{flucM}) as
\bea
m_{AB} &\equiv&
\left(
\begin{array}{cc}
a & -b \\
b & -a
\end{array}
\right)_{AB}\;,
\label{linS}
\eea
with symmetric $a$  and antisymmetric $b$, 
in accordance with the $SO(d,d)$ property of $M_{AB}$,
we finally obtain the fluctuations
\bea
\begin{split}
\hat{h}_{mn}+\hat{k}_{mn}&=S^{ab}(x)L_{a\,n}(y)R_{b\,m}(y)\,,\\
\phi &=\varphi(x)+\frac{1}{4}S^{ab}(x)L_{a}{}^m(y)R_{b\,m}(y)\,,
\end{split}
\label{linhk}
\eea
with $S^{ab}\equiv 4 \left(a^{ab}+b^{ab}\right)$. 
These precisely reproduces the linearised result given in \cite{Duff:1986ya}.

After the full non-linear reduction (\ref{reducG})--(\ref{reducform}),
the reduced theory is an $n$-dimensional gravity coupled to a 2-form and $2d$ gauge vectors with
gauge group $G_L\times G_R$.
The $(d^2+1)$ scalar fields couple as an $\mathbb{R} \times SO(d,d)/(SO(d)\times SO(d))$
coset space sigma model, 
and come with a scalar potential~\cite{Kaloper:1999yr,Schon:2006kz}
\bea
V(x) &=& \frac1{12}\,e^{4\gamma\varphi(x)}
X_{AB}{}^{C} X_{DE}{}^{F} M^{AD}(x) \left( M^{BE}(x)M_{CF}(x) + 
3\,\delta_C^E\delta_F^B\, \right)
\;,
\label{Vpot}
\eea
with the structure constants $X_{AB}{}^{C}$ from (\ref{XABC}).
Due to the dilaton prefactor, this potential cannot support (A)dS geometries,
but only Minkowski or domain wall solutions.

Let us finally comment on adding a cosmological term $e^{4\beta\phi}\Lambda$ in the higher-dimensional theory \eqref{string}. 
E.g.\ for the bosonic string such a term would arise as conformal anomaly in dimension $n+d\not=26$\,.
In the Einstein frame, the modified action takes the form
\bea
 S &=& \int dX^{n+d}\,\sqrt{|G|}\Big(  {R}
   +4\,G^{\hat\mu\hat\nu}\partial_{\hat\mu}\phi \partial_{\hat\nu}\phi 
    -\frac{1}{12}\,e^{-8\beta\phi}\,{H}^{\hat\mu\hat\nu\hat\rho}{H}_{\hat\mu\hat\nu\hat\rho}+e^{4\beta\phi}\Lambda
 \Big)\;,
 \label{stringEin}
\eea 
with constant $\Lambda$. With the $O(d,d)$ dictionary (\ref{dictionary2}), it follows that the effect of this term
in the $O(d,d)$ covariant action (\ref{DFTAction}) is a similar term
\bea
{\cal L}_{c} &=&\sqrt{|{\rm g}|}\,e^{-2\Phi}\Lambda\;,
\label{ccDFT}
\eea
manifestly respecting $O(d,d)$ covariance. The presence of this term thus does not interfere with the consistency of 
the truncation ansatz and simply results in a term
\bea
{\cal L}_{c} &=& \sqrt{|g|}\,e^{4\gamma\varphi}\Lambda\;,
\label{Vcc}
\eea
in the reduced theory, as already argued in \cite{Duff:1986ya,Cvetic:2000dm}.

\section{Conclusions}

We have in this paper given a complete and constructive proof of the consistency of the Pauli reduction 
of the low-energy effective action of the bosonic string on the group manifold $G$,
proving the conjecture of~\cite{Duff:1986ya}.
The construction is based on the $O(d,d)$ covariant reformulation of the original theory
in which the consistent truncations of the latter are rephrased as generalised 
Scherk-Schwarz reductions on an extended spacetime.
We have explicitly constructed the relevant $SO(d,d)$ valued twist matrix, 
carrying the dependence on the internal variables, in terms of the Killing vectors 
of the group manifold $G$. From the twist matrix, we have further read off the 
full non-linear reduction ans\"atze  for all fields of the bosonic string. 
The construction is another example of the power of the generalised 
Scherk-Schwarz reductions on extended spacetime and hints 
towards a more systematic understanding of the conditions under which 
consistent Pauli reductions are possible.
In this respect, it would be very interesting to classify the possible solutions of the 
system of equations~(\ref{system}) encoding the consistent reduction.

For a compact group manifold $G$, the obtained twist matrix describes the
consistent Pauli reduction of the bosonic string on $G$.
Interestingly, the construction straightforwardly generalises to the case 
when $G$ is a non-compact group. In this case, the resulting twist matrix
is still built from the Killing vectors on $G$, but
describes the consistent reduction of the bosonic string on an internal 
manifold $M_d$ whose metric is read off from (\ref{Ginv}) as
\bea
\mathring{G}^{mn} &=& ({\rm det\,}\mathring{H}/{\rm det\,}
\tilde{G})^{\beta}  \, \mathring{H}^{mn}
\;,\qquad\mbox{with}\;\;
\mathring{H}^{mn} ~\equiv~ \cK_A{}^m(y)\cK_B{}^n(y) \,\delta^{AB}
\;,
\label{Gnc}
\eea
and $\tilde{G}_{mn}$ defined in (\ref{Gtilde}) as the bi-invariant metric on $G$.
It follows that the isometry group of this background metric $\mathring{G}_{mn}$
is the maximally compact subgroup $K_{L}\times K_{R}\subset G_L\times G_R$.
The truncation in this case retains not only the gauge bosons of the isometry group, but
the gauge group of the lower-dimensional theory enhances to the full non-compact $G_L\times G_R$.
At the scalar origin, the non-compact gauge group is broken down to its compact part $K_{L}\times K_{R}$.
This is a standard scenario in supergravity.
For the known sphere reductions the corresponding generalisation describes the
compactification on non-compact hyperboloidal spaces $H^{p,q}$ inducing lower-dimensional theories with $SO(p,q)$
gauge groups~\cite{Hull:1988jw,Cvetic:2004km,Hohm:2014qga,Baron:2014bya}.
Similar to (\ref{Gnc}), the background 2-form $\mathring{C}_{mn}$ is read off from (\ref{reducform}) 
and in this case differs from $\tilde{C}_{mn}$ by a contribution from the second term.

In general, the background geometry (\ref{background_cp}) or (\ref{Gnc}) does not provide a solution to the
higher-dimensional field equations. This corresponds to the fact that the scalar potential (\ref{Vpot}) in
general does not possess a stationary point at the scalar origin.
However, a quick computation shows that at the origin  $M_{AB}=\delta_{AB}$, the potential (\ref{Vpot}) is 
always stationary with respect to 
variation of the parameters of $M_{AB}$, such that there is a ground state with 
running dilaton $\varphi(x)$\,. Via (\ref{reducG})--(\ref{reducform}), this domain wall solution is
uplifted to the higher-dimensional theory.

A necessary and sufficient condition for the existence of a ground state with constant dilaton 
is $V|_{M_{AB}=\delta_{AB}}=0$, i.e.\ necessarily a Minkowski vacuum. Evaluating the scalar potential
at the origin translates this condition into
\bea
0~\stackrel{!}{=}~ V\Big|_{M_{AB}=\delta_{AB}} &=& \frac23\left(2\,n_{\mbox{\scriptsize non-cp}}-n_{\rm cp} \right)
\;,
\label{Minkowski}
\eea
with $n_{\rm cp}$, $n_{\rm non-cp}$ denoting the number of compact and non-compact
generators of $G$, respectively. 
A number of groups satisfy this condition
\bea
G&=& SO(1,5)\;,\;\; SO(5,20)\;,\;\; SO(20,76)\;,\;\;\dots
\;,
\nonumber\\
G&=& SU(1,4)\;,\;\; SU(4,15)\;,\;\; SU(15,56)\;,\;\;\dots
\;,
\nonumber\\
G &=& E_{6(-26)}\;,\;
\mbox{with compact $F_4$}\;,
\eea
thus allowing for a consistent truncation of the bosonic string around ${\rm Mink}_{n} \times M_{{\rm dim}\,G}$,
the latter equipped with the metric (\ref{Gnc}).
While these examples are presumably more of a mathematical curiosity, a group of more physical relevance
is the choice
\bea
G&=& SO^*(4)~\equiv~ SO(3)\times SO(2,1)
\;,
\label{G4}
\eea
satisfying the condition (\ref{Minkowski}). With this group, the above construction  
describes the consistent truncation of ten-dimensional $N=1$ supergravity down
to four dimensions on the manifold $S^3 \times H^{2,2}$ giving rise to
a half-maximal $SO(4)\times SO(2,2)$ gauged theory in four dimensions with Minkowski vacuum.
The form of the vacuum resembles the Minkowski vacua found in \cite{Pernici:1984nw}
with uplift to eleven dimensions.
It would be very interesting to embed this vacuum into the maximal theories 
allowing for Minkowski vacua~\cite{DallAgata:2011aa,Catino:2013ppa}
whose respective gauge groups $SO^*(8)$ and $SO(4)\times SO(2,2)\ltimes T^{16}$ indeed 
contain two copies of (\ref{G4}).
The embedding of the maximal theory in higher dimensions may then be addressed similar to
the construction of this paper
within the proper full exceptional field theory~\cite{Hohm:2013uia}.

Among the interesting examples with running dilaton, our construction includes the consistent truncation
on $S^3\times S^3$ corresponding to the compact choice $G=SO(4)$. 
In this case, the above construction gives the consistent embedding
of $SO(4)^2$-gauged half-maximal supergravity into ten dimensions, extending the
construction of~\cite{Chamseddine:1997mc}, in which the scalar sector was
truncated to the dilaton. Again, it would be interesting to embed this truncation into
the maximal theory. It is likely that different embeddings into IIA and IIB may give rise to
inequivalent maximal four-dimensional gaugings, as observed for the IIA/IIB $S^3$ reductions 
to seven dimensions~\cite{Malek:2015hma}.

Let us finally mention that the presented construction provides another 
example of globally geometric non-toroidal compactifications inducing non-geometric fluxes. 
In the language of~\cite{Shelton:2005cf}, the structure constants~(\ref{XABC}) induced by this reduction
combine a 3-form flux $H_{abc}$ with non-geometric $Q_a{}^{bc}$ flux. 
However, despite their non-geometric appearance, the fluxes satisfy the condition $f_{KMN}f^{KMN}=0$, 
necessary for a potential geometric origin~\cite{Dibitetto:2012rk}, which we have provided here. 
For the compactifications on $S^3$ and $S^3\times S^3$, this scenario has been
discussed in \cite{Danielsson:2015tsa,Danielsson:2015rca}.

\section*{Acknowledgements}
We thank Gianguido Dall'Agata and Olaf Hohm
for helpful discussions.
C.N.P. is supported in part by DOE grant DE-FG02-13ER42020.


\providecommand{\href}[2]{#2}\begingroup\raggedright\endgroup

\end{document}